\documentclass{emulateapj}
\newcommand{\rev}[1]{{#1}}

\begin{document}

\title{From Planetesimals to Planets in Turbulent Protoplanetary Disks I. \\
Onset of Runaway Growth 
}

\shorttitle{
Onset of Runaway Growth in Turbulent Disks 
}
\shortauthors{
Kobayashi, Tanaka, Okuzumi
}

\author{Hiroshi Kobayashi\altaffilmark{1}
Hidekazu Tanaka\altaffilmark{2} and 
Satoshi Okuzumi\altaffilmark{3}
} 

\altaffiltext{1}{
Department of Physics, Nagoya University, Nagoya, Aichi 464-8602, Japan
}
\email{hkobayas@nagoya-u.jp}

\altaffiltext{2}{
Institute of Low Temperature Science, Hokkaido University,
Kita-Ku Kita 19 Nishi 8, Sapporo 060-0819, Japan
}
\email{hide@lowtem.hokudai.ac.jp}

\altaffiltext{3}{Tokyo Institute of Technology, Ookayama, Meguro-ku, Tokyo 152-8551, Japan}
\email{okuzumi@geo.titech.ac.jp}

\begin{abstract}
When planetesimals grow via collisions in a turbulent disk, stirring
 through density fluctuation caused by turbulence effectively increases
 the relative velocities between planetesimals, which suppresses the
 onset of runaway growth.  We investigate the onset of runaway growth in
 a turbulent disk through simulations that calculate the mass and
 velocity evolution of planetesimals.  When planetesimals are small, the
 average relative velocity between planetesimals, $v_{\rm r}$, is much
 greater than their surface escape velocity, $v_{\rm esc}$, so that
 runaway growth does not occur.  As planetesimals become large via
 collisional growth, $v_{\rm r}$ approaches $v_{\rm esc}$.  When $v_{\rm
 r} \approx 1.5 v_{\rm esc}$, runaway growth of the planetesimals
 occurs. During the oligarchic growth subsequent to runaway growth, a
 small number of planetary embryos produced via runaway growth become
 massive through collisions with planetesimals with radii of that at the
 onset of runaway growth, $r_{\rm p,run}$.  We analytically derive
 $r_{\rm p,run}$ as a function of the turbulent strength. Growing $\sim
 10\,M_\oplus$ embryos that are suitable to become the cores of Jupiter
 and Saturn requires $r_{\rm p,run} \sim 100$\,km, which is similar to
 the proposed fossil feature in the size distribution of main belt
 asteroids. In contrast, the formation of Mars as quickly as suggested
 from Hf-W isotope studies requires small planetesimals at the onset of
 runaway growth.  Thus, the conditions required to form Mars, Jupiter, and
 Saturn and the size distribution of the main-belt asteroids indicate
 that the turbulence increased in amplitude relative to the sound speed
 with increasing distance from the young Sun.
\end{abstract}


\vspace{1cm}

\section{INTRODUCTION}
\label{sc:intro}

Planets are considered to be formed in a protoplanetary disk composed of
gases and solids. In the standard scenario, kilometer-sized or larger planetesimals 
are generated from dust grains and collisional
coagulation of the planetesimals forms planetary embryos. Once planetary
embryos are as large as 10 Earth masses, the embryos start rapid gas
accretion, which results in gas giant planets 
\citep{ikoma00,hori10,mizuno80}. 

A swarm of planetesimals produces planetary embryos through runaway
growth \citep{wetherill89,kokubo96,ormel10} and the embryos grow further
through the accretion of remnant planetesimals, which is called
oligarchic growth \citep{weidenschilling97,kokubo98}. However, the
stirring by embryos is so strong that collisions between planetesimals
are destructive. The fragments produced by planetesimal collisions become progressively
smaller by collisional cascade until 10 meter-sized
fragments drift inward by gas drag \citep{kobayashi+10}. The collisional
fragmentation of planetesimals and the radial drift of small bodies
that results from collisional fragmentation reduce the solid surface
density of bodies surrounding planetary embryos, so that the growth of
the planetary embryos is stalled \citep{kobayashi+10,kobayashi+11}.  The collisional outcomes for
kilometer-sized or larger planetesimals are determined by re-accretion
of collisional fragments that result from shattering; therefore, the
effective collisional strength is mainly controlled by self-gravity, so that
larger planetesimals are collisionally stronger.  Larger planetesimals
can avoid a reduction of the solid surface density due to
collisional fragmentation, which results in the formation of more massive planetary embryos.
On the other hand, the timescale for planetary embryo formation through
runaway and oligarchic growth is longer for larger
planetesimals. Therefore, \citet{kobayashi+11} found that the formation
of 10~Earth mass cores inside 10\,AU within 10\,Myr requires
moderate-sized planetesimals at the onset of runaway growth ($\sim 100$\,km in radius) in a massive disk ($\sim 0.1$ solar masses).

Dust particles grow to meter-sized pebbles via collisions in a
protoplanetary disk.  Pebbles lose substantial angular momentum due to
gas drag because of the sub-Keplerian rotational velocity of gas and 
spiral onto the central star. If pebbles are compact, then radial drift is too
rapid to grow to planetesimals \citep{weidenschilling77}. However,
successive collisions of dust produce highly porous aggregates
\citep{suyama08,suyama12}. Icy porous aggregates \rev{that consist of a
number of spherical sub-micron particles} do not suffer from
catastrophic disruption unless the impact velocities exceed 60--90\,m/s
\citep{wada09,wada13}.  \citet{okuzumi12} investigated the collisional
evolution of the mass and porosity of icy dust aggregates, whereby icy fluffy dust
aggregates were determined to grow into planetesimals faster than they drift. The filling
factor of the aggregates becomes as low as $\sim 10^{-5}$ when radial
drift is most effective. \rev{Since the compaction of aggregates by 
ram pressure and self-gravity is effective with the growth of aggregates, the filling factor of kilometer-sized
or larger planetesimals increases up to $\sim 0.1$ \citep{kataoka13a,kataoka13b}.}

The critical collisional velocity that inhibits collisional growth is
$80$\,m/s for icy dust \rev{aggregates} \citep{wada13}. Such a
high critical velocity means that icy planetesimals can be formed through
collisions.  Silicate
dust aggregates may be destroyed at $8$\,m/s if the surface
energy\footnote{\rev{The adhesion energy between particles is
proportional to $\gamma^{5/3}$ \citep{JKR}. The critical fragmentation 
velocity of aggregates composed of the particles is proportional to the square root of the
adhesion energy (e.g., $\propto \gamma^{5/6}$) \citep{wada09,wada13}.}}
$\gamma$ of the
particles is $0.025\,{\rm J m}^{-2}$ \citep{wada13}. However,
\citet{kimura} pointed out that the surface energy of silicates
depends on the outermost layer of absorbed water on the surfaces of dust particles. Thus, for
a small amount of absorbed water, $\gamma$ can be larger than $0.1 \,
{\rm J m}^{-2}$ \citep{wiederhorn,shchipalov,han,roder,tromans}. 
If $\gamma = 0.25
\,{\rm J m}^{-2}$, then the critical collisional velocity of silicate
aggregates can be larger than $50\,{\rm m/s}$ (Kimura et al. 2015),
which allows the formation of silicate planetesimals via collisional growth. 
\rev{Therefore, here we consider planetesimal formation via the collisional
coagulation of dust aggregates, while we ignore the alternative paths for
planetesimal formation, such as streaming instability \citep{johansen}. }

The collisional growth of porous aggregates overcomes the drift barrier and 
the subsequent compaction of aggregates produces planetesimals with low porosity,
which then grow further via collisions. Once gravitational focusing and
dynamical friction are effective, the runaway growth of planetesimals occurs \citep{wetherill89}. 
Runaway growth produces a small number of planetary embryos, which
then grow further via collisions with surrounding planetesimals that have radii 
similar to that \rev{at the onset of} runaway growth. The lifetime of the
surrounding planetesimals due to 
collisional fragmentation depends on their size
\citep{kobayashi10,kobayashi+10}. Thus, 
the size of planetesimals at the onset of runaway growth affects the
growth of planetary embryos. 

Gravitational focusing and
dynamical friction are effective if $v_{\rm r} \la v_{\rm esc}$, where
$v_{\rm r}$ is the relative 
velocity between planetesimals and $v_{\rm esc}$ 
is the mutual surface escape velocity of the planetesimals; however, runaway growth
is suppressed if $v_{\rm r} \ga v_{\rm esc}$ \citep{wetherill89}. 
In protoplanetary disks, turbulent stirring due 
to hydrodynamical gas drag accelerates $v_{\rm r}$, which is effective
for bodies smaller than $\sim 1\,$m. 
On the other hand, the density fluctuation caused by turbulence
results in significant perturbation that effectively increases the random velocities of
kilometer-sized or larger planetesimals \citep{ida,okuzumi13,ormel13}. 
Stirring by the density 
fluctuation in a turbulent disk suppresses the runaway growth of
such kilometer-sized or larger planetesimals. 
Once runaway growth occurs, the planetary embryos formed are surrounded by a
swarm of planetesimals with radii similar to that at the onset of runaway
growth, \rev{which is similar to the condition resulting from previous
studies starting from planetesimals \citep[e.g.,][]{kokubo96,kokubo98,inaba01}}. Therefore, if \rev{representative planetesimals have radii of $\sim100$\,km at the onset of} runaway growth, as suggested in
\citet{kobayashi+11}, then planetesimals may form massive cores to become gas
giants. 

Here, we investigate the collisional evolution of planetesimals by
taking turbulent stirring into account. In Section \ref{sc:model}, we
introduce a model for turbulent stirring, collisional evolution, and
protoplanetary disks. In Section \ref{sc:simulation}, 
simulations are conducted for the collisional evolution of planetesimals in turbulent
disks. In Section \ref{sc:analy}, the average
relative velocities of planetesimals in a turbulent disk are analytically derived and the
radius of the planetesimals \rev{at the onset of} runaway growth is then determined.  In Section
\ref{sc:subsequent}, we discuss the radial profile of the turbulence strength
required to form the Solar System.  We summarize our conclusions in Section \ref{sc:discussion}.  In a separate paper (Paper II), 
continuous simulations are performed from the stage prior to runaway growth until the
formation and growth of planetary embryos.

\section{Modeling for Collisional Growth in a Turbulent Disk}
\label{sc:model}

\subsection{Collisional Growth}

Collisions between bodies at a distance $a$ from the host star
\rev{and the radial drift of bodies}
evolve the surface number density of bodies with 
masses ranging from $m$ to $m+dm$, $n_{\rm s}(m,a) dm$, as
\begin{eqnarray}
 \frac{\partial m n_{\rm s}(m,a)}{\partial t} &=& \frac{m}{2} \int_0^\infty
  dm_1 \int_0^\infty dm_2 
\nonumber\label{eq:coag}
\\
  && \quad \times n_{\rm s}(m_1,a) n_{\rm s}(m_2,a) K(m_1,m_2)
\nonumber 
\\
  && \quad \times \delta(m-m_1-m_2+m_{\rm e})
\nonumber
\\
&& - m n_{\rm s}(m) \int_0^\infty dm_2 n_{\rm s} (m_2,a) K(m,m_2) 
\nonumber
\\
&& + \frac{\partial}{\partial m} \int_0^\infty dm_1 \int_0^{m_1 } dm_2 
n_{\rm s}(m_1,a) n_{\rm s}(m_2,a) 
\nonumber
\\
 && \quad \times K(m_1,m_2) \Psi(m,m_1,m_2)
\nonumber
\\
&& - \frac{1}{a} \frac{\partial}{\partial a} [ a m n_{\rm s}(m,a) v_{\rm
 drift}(m,a)], 
\end{eqnarray}
where $m_{\rm e}$ and 
$\Psi(m,m_1,m_2)$ are, respectively, the total and cumulative masses of
bodies ejected by a single collision between bodies with masses $m_1$
and $m_2$, $v_{\rm drift}$ is the drift velocity of a body due to
gas drag, and 
\begin{equation}
 K(m_1,m_2) = (h_{m_1,m_2} a)^2 \langle {\cal P}_{\rm col}(m_1,m_2)
  \rangle \Omega
\end{equation}
where $h_{m_1,m_2} = [(m_1+m_2)/3M_*]^{1/3}$ is the reduced mutual Hill
radius, $\langle {\cal P}_{\rm col}(m_1,m_2) \rangle$ is the dimensionless
mean collisional rate, and $\Omega$ is the Keplerian frequency. The collisional rate $\langle {\cal P}_{\rm col}(m_1,m_2)
\rangle$ is given by a function of eccentricities and inclinations of
colliding bodies, as summarized in \citet{inaba01}.

The relative velocities between bodies are determined by their
eccentricities and inclinations. 
The dispersions for eccentricities and inclinations change according to
the gravitational interaction between bodies \citep{ohtsuki02}, damping
by gas drag \citep{adachi76}, and collisional damping
\citep{ohtsuki92}. Stirring by turbulence is additionally taken into account and is introduced below. The eccentricity and inclination evolution is
expressed as 
\begin{eqnarray}
 \frac{d e^2}{dt} &=&
  \left. \frac{d e^2}{dt} \right|_{\rm grav}
 + \left. \frac{d e^2}{dt} \right|_{\rm drag}
 + \left. \frac{d e^2}{dt} \right|_{\rm coll}
 + \left. \frac{d e^2}{dt} \right|_{\rm turb},\label{eq:de_sim}\\
\frac{d i^2}{dt}  &=&
  \left. \frac{d i^2}{dt} \right|_{\rm grav}
 + \left. \frac{d i^2}{dt} \right|_{\rm drag}
 + \left. \frac{d i^2}{dt} \right|_{\rm coll}
 + \left. \frac{d i^2}{dt} \right|_{\rm turb},\label{eq:di_sim}
\end{eqnarray}
where the subscripts ``grav'', ``drag'', ``coll'', and ``turb'' indicate 
gravitational interaction, gas drag damping, collisional damping, and
turbulent stirring, respectively.  The damping rates by gas drag
are given functions composed of the lowest order terms of $e$ and $i$ following \citet{inaba01} because $e$ and $i$
are much smaller than unity, although the damping rates by gas drag have
higher order terms for high $e$ and $i$ as shown in
\citet{kobayashi15}.

Turbulence stirring increases the random velocities of planetesimals, which
delays the onset of runaway growth. Collisional destruction is less
important before and during runaway growth in the disk without
turbulence \citep{kobayashi+10}. Therefore, collisional
fragmentation is ignored by setting $m_{\rm e} = 0$ and $\Psi(m,m_1,m_2) = 0$ in
Eq.~(\ref{eq:coag}); however, this effect is investigated in Paper II.

For the initial condition, the respective surface densities of gas and solid are
given by
\begin{eqnarray}
 \Sigma_{\rm g} &=& 1.7 \times 10^3 x_{\rm g}
  \left(\frac{a}{1\,{\rm AU}}\right)^{-1.5} \, {\rm g \, cm}^{-2}\label{eq:sigmag}\\
 \Sigma_{\rm s} &=& 30 f_{\rm ice} x_{\rm s} \left(\frac{a}{1\,{\rm AU}}\right)^{-1.5}
  \, {\rm g \, cm}^{-2},\label{eq:sigmas}   
\end{eqnarray}
where the ice factor $f_{\rm ice}$ is given as unity beyond the snow
line\rev{,} $x_{\rm g}$ and $x_{\rm s}$ are scaling factors\rev{,} and the disk
with $x_{\rm g} = x_{\rm s} =1$ corresponds to the minimum-mass solar
nebula (MMSN) model \citep{hayashi}. We consider disks beyond the snow
line ($f_{\rm ice} = 1$) around solar type stars with \rev{$M_* =
M_\sun$}. Disks ranging from 4.3 to 67\,AU are treated using 8 radial
meshes\rev{. Bodies initially have radii of
 1\,km and $e = 10 i = 6 \times 10^{-4}$. As
analytically shown in \S~\ref{sc:analy}, the results are almost
independent of the initial size distribution, $e$, and $i$ of the bodies}.  The solid surface
density varies through collisional evolution and radial draft, whereas, for simplicity,
the gas surface density does not change.

\subsection{Turbulent Stirring}

The turbulent stirring is given by 
\begin{eqnarray}
 \left. \frac{d e^2}{dt} \right|_{\rm turb} &=& 
 \left. \frac{d e^2}{dt} \right|_{\rm std} + 
 \left. \frac{d e^2}{dt} \right|_{\rm fs},\\
 \left. \frac{d i^2}{dt} \right|_{\rm turb} &=& 
 \left. \frac{d i^2}{dt} \right|_{\rm std} + 
 \left. \frac{d i^2}{dt} \right|_{\rm fs},
\end{eqnarray}
where the subscripts ``std'' and ``fs'' indicate density fluctuation
stirring and frictional stirring, respectively. Each term is explained
below. 

Magnetorotational instability (MRI) induces turbulence in
protoplanetary disks \citep[e.g.,][]{balbus91,suzuki10}. 
The excitation rate of eccentricities due to the density fluctuation caused by
MRI turbulence is given by \citep{okuzumi13}
\begin{equation}
 \left. \frac{de^2}{dt} \right|_{\rm sdf}= f_{\rm d}   
  \left(\frac{\Sigma_{\rm g} a^2}{M_*}\right)^2 \Omega\label{eq:sdf} 
\end{equation}
where $f_{\rm d}$
is the dimensionless factor dependent on the strength of
the density fluctuation. \citet{okuzumi13} compiled previous results of
magnetohydrodynamical simulations for MRI turbulent disks
\citep{okuzumi11,gressel} and semi-analytically
obtained $f_{\rm d}$, given by 
\begin{equation}
f_{\rm d}  = \frac{0.94 {\cal L} \alpha}{(1+4.5 H_{\rm
 res,0}/H)^2}, 
\end{equation}
where $H$ is the scale height of the disk, $H_{\rm res,0}$ is the
half vertical width of the dead zone, $\alpha$ is the dimensionless
viscosity at the midplane of the disk scaled by the sound velocity and the scale
height, and ${\cal L}$ is a scale parameter of the order of unity. In this
paper, ${\cal L} = 1$ for simplicity. 

Orbital inclination is also increased due to the density fluctuation. However, 
the excitation rate of inclination is much smaller than that of eccentricity in MRI turbulence. We assume 
\begin{equation}
  \left. \frac{di^2}{dt} \right|_{\rm sdf}= \epsilon^2
 \left. \frac{de^2}{dt} \right|_{\rm sdf},\label{eq:di_sdf} 
\end{equation}
where the coefficient $\epsilon$ is much smaller than unity and of the
order of 0.1 \citep{yang12}. $\epsilon$ is set as $0.1$. 

The other turbulent stirring is caused by aerodynamical friction force \citep{voelk80}. If the stopping time by gas drag is longer
than the orbital period, then the stirring rates for random velocities
are 
given by \citet{youdin07} under the assumption of isotropic turbulence. The stirring rate of eccentricity is
twice as large as that of inclination due to isotropic turbulence. Using the stirring rate given by
\citet{youdin07} under this assumption, the stirring rates are then
approximated to be 
\begin{eqnarray}
 \left. \frac{de^2}{dt} \right|_{\rm fs} &=& \frac{4 \alpha}{3 \Omega
  t_{\rm stop}^2 } 
\left(\frac{c_{\rm s}}{\Omega a} \right)^2,\label{eq:de_fs} \\
 \left. \frac{di^2}{dt} \right|_{\rm fs} &=& 
\frac{2 \alpha}{3 \Omega
  t_{\rm stop}^2 }
\left(\frac{c_{\rm s}}{\Omega a} \right)^2,\label{eq:di_fs}  
\end{eqnarray}
where $t_{\rm stop}$ is the stopping time and 
$t_{\rm stop}
\gg \Omega^{-1}$ is assumed. \rev{For a body with mass $m$
and radius $r_{\rm p}$, 
\begin{equation}
 t_{\rm stop} = \frac{2m}{\pi C_{\rm D} r_{\rm p}^2 \rho_{\rm gas} u}, 
\end{equation}
where $\rho_{\rm gas}$ is the gas density, $u$ is the relative velocity, and
$C_{\rm D}$ is the dimensionless drag coefficient. Although $C_{\rm D}$
is given by a function of the Mach and Reynolds numbers in 
simulations according to \citet{kobayashi+10}, $C_{\rm D} \approx 0.5$
because bodies with $r_{\rm p} > 1\,$km are treated in this paper. }

Turbulent stirring via friction is important only for bodies $\la
100\,$m. Although this stirring has little effect on the onset of
runaway growth, it is included for consistency. 

\section{Evolution of Size Distribution and Random Velocities}
\label{sc:simulation}

\begin{figure}[htbp]
\epsscale{0.99} \plotone{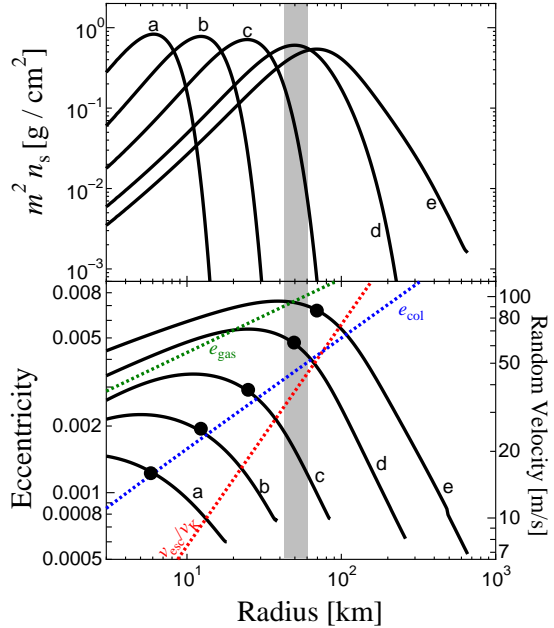} \figcaption{ Surface
density of planetesimals $m^2 n_{s}$ (top), and eccentricity (bottom) at (a)
$1.3 \times 10^5$\,yr, (b) $3.6 \times 10^5$\,yr, (c) $8.9 \times
10^5$\,yr, (d) $2.3 \times 10^6$\,yr, and (e) $3.7 \times 10^6$\,yr for
$x_{\rm g} = x_{\rm g} = 1$ (MMSN), $\rho_{\rm s} = 1\,{\rm g/cm}^3$,
$\alpha = 10^{-2}$, and $H_{\rm res,0} = H$ ($f_{\rm d} \approx 3.1 \times
10^{-4}$) as a function of the radii of planetesimals.  The runaway growth of
planetesimals starts at the grey hatched area.  Filled circles indicate
eccentricities at the peak of $m^2 n_{\rm s}$.  The red, blue, and green
dotted lines indicate $v_{\rm esc}/v_{\rm K}$, $e_{\rm col}$
(Eq.~\ref{eq:ecol}), and $e_{\rm gas}$ (Eq.~\ref{eq:egas}),
respectively.  
\label{fig:MMSN}
} 
\end{figure}

Logarithmic mass bins are set with width $\delta m = 1.05 m$, and the
method developed by \citet{kobayashi+10} is used to simultaneously integrate
Eqs.~(\ref{eq:coag}), (\ref{eq:de_sim}), and (\ref{eq:di_sim}) to derive
the time evolution of the size distribution, eccentricities, and
inclinations.  Fig.~\ref{fig:MMSN} shows $m^2 n_{\rm s}$ and $e$ at
5.2\,AU for $x_{\rm s} = x_{\rm g} =1$, $f_{\rm d} \approx 3.1 \times
10^{-4}$ ($\alpha = 0.01$ and $H_{\rm res,0} = H$), and the internal
density of bodies of $\rho_{\rm s} = 1 {\rm g/cm}^3$.  The solid surface
density is given by $\Sigma_{\rm s} \equiv \int m n_{\rm s} dm = \int
m^2 n_{\rm s} d \ln m$; hence $m^2 n_{\rm s}$ indicates the surface
density of planetesimals with masses around $m$.  The planetesimal
radius at the peak of $m^2 n_{\rm s}$ is given approximately by the
weighted average radius of planetesimals, defined as
\begin{equation}
 {\bar r}_{\rm p} \equiv \frac{1}{\Sigma_{\rm s}} \int r_{\rm p} m
  n_{\rm s} d m, 
\end{equation}
where $r_{\rm p}$ is the radius of a planetesimal with mass $m$.  At the
early stage ($t \ll 2 \times 10^6$\,yr), $m^2 n_{\rm s}$ has a narrow
peak around ${\bar r}_{\rm p}$; most planetesimals have radii similar to 
${\bar r}_{\rm p}$. This indicates the orderly growth of
planetesimals.  The shapes around the peak of $m^2 n_{\rm s}$ are
similar over time. However, at $\approx 2\times10^6$\,yr, the runaway
growth of planetesimals makes the peak wider (see Fig.~\ref{fig:MMSN}).
Note that the peak of the planetesimal size distribution is almost
unchanged after runaway growth starts. 

The eccentricities of planetesimals
depend on the size of planetesimals (see Fig.~\ref{fig:MMSN}). At $r_{\rm p} = {\bar r}_{\rm p}$,
$e$ is approximately equal to $e_{\rm col}$ determined by the
turbulent stirring, and the collisional growth and damping (see
derivation in Section \ref{sc:velocity}).  For bodies larger than ${\bar
r}_{\rm p}$, the dynamical friction between bodies controls $e$ if $e <
v_{\rm esc}/v_{\rm K}$, where $v_{\rm K} = a \Omega$ is the Keplerian
velocity. For bodies smaller than ${\bar r}_{\rm p}$, $e$ is larger than
$e_{\rm col}$ because collisional damping has little effect on $e$ due
to less frequent collisions between the bodies. The onset of runaway
growth is determined by $v_{\rm r} \approx e v_{\rm K}$ of planetesimals
with ${\bar r}_{\rm p}$. When ${\bar r}_{\rm p}$ is small, $e$ for the
planetesimals with ${\bar r}_{\rm p}$ is much larger than $v_{\rm
esc}/v_{\rm K}$.  Once $e$ for the planetesimals with ${\bar r}_{\rm p}$
is close to $v_{\rm esc}/v_{\rm k}$, gravitational focusing is no
longer negligible, which ignites the runaway growth of
planetesimals. Runaway growth occurs at 40--60\,km in radius. After
the onset of runaway growth, the largest planetesimals grow faster than
the planetesimals with weighted average radius. The stirring of the largest
planetesimals that result from the runaway growth increases the eccentricities of
planetesimals with ${\bar r}_{\rm p}$.

\begin{figure}[htbp]
\epsscale{0.9}
\plotone{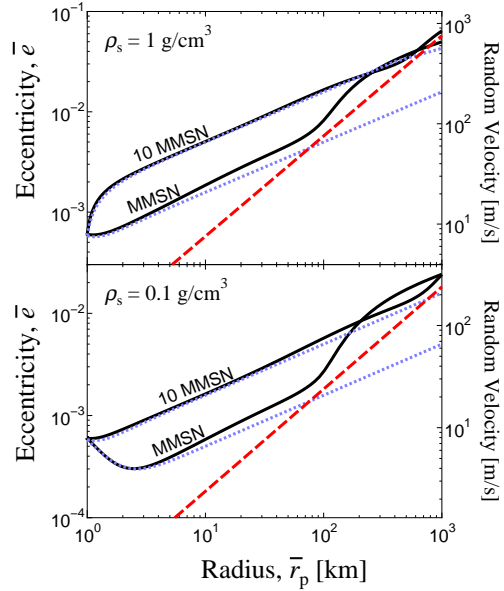}
\figcaption{
Relation between the weighted average radius and the weighted average
 eccentricity (black curves) during collisional
 growth for $\rho_{\rm s} = 1\,{\rm g/cm}^3$ (top panel) and $\rho_{\rm
 s} = 0.1 \,{\rm g/cm}^3$ (bottom panel), 
$r_{\rm p0} = 1$\,km, $f_{\rm d} \approx 3 \times 10^{-4}$
($\alpha = 10^{-2}$ and $H_{\rm res,0} = H$) 
at 5.2\,AU 
in disks with $x_{\rm g} = x_{\rm s} = 1$ (MMSN; bottom curves
 around 1\,km) and
 10 (10~MMSN; top curves around 1\,km). 
Dotted curves (blue) indicate the smaller of $e_{\rm col}$ and $e_{\rm
 gas}$ given in Eqs. (\ref{eq:ecol}) and (\ref{eq:egas}),
 respectively. Dashed lines (red) are the surface escape velocity
 divided by the Keplerian velocity for reference.  
\label{fig:ecc_radius}
}
\end{figure}

Fig.~\ref{fig:ecc_radius} shows the evolution of ${\bar r}_{\rm p}$ and
${\bar e}$ in the simulations (solid curves), where ${\bar e}$ is the
weighted average eccentricity, defined as
\begin{equation}
 {\bar e} \equiv \frac{1}{\Sigma_{\rm s}} \int e m n_{\rm s} dm. 
\end{equation}
As ${\bar r}_{\rm p}$ increases, ${\bar e}$ approaches $v_{\rm
esc}/v_{\rm K}$.  The runaway growth of
planetesimals occurs at ${\bar r}_{\rm p} \approx 40$--60\,km for
$f_{\rm d} \approx 3.1 \times 10^{-4}$ and $\rho_{\rm s} = 1\,{\rm
g/cm}^3$ in the MMSN disk, as shown in Fig.~\ref{fig:MMSN}. When ${\bar r}_{\rm p}$ becomes 
$40$--60\,km, ${\bar e}$ is comparable to $v_{\rm esc}/v_{\rm K}$ (see
Fig.~\ref{fig:ecc_radius}). For ${\bar r}_{\rm p} \ga 90$\,km, ${\bar
e}$ rapidly increases with ${\bar r}_{\rm p}$.  The runaway
growth of planetesimals rapidly produces planetesimals much larger than
planetesimals with the weighted average radius.  
Fig.~\ref{fig:MMSN} also shows that stirring by the largest planetesimals produced by
runaway growth increases the eccentricity of planetesimals with ${\bar
r}_{\rm p}$.

\begin{figure}[htbp]
\epsscale{0.99}
\plotone{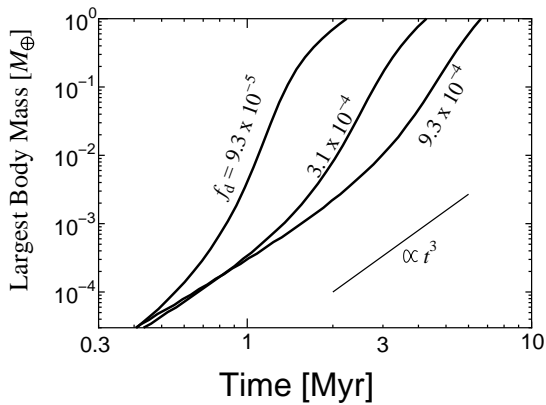}
\figcaption{
Time evolution of the mass of the largest planetesimals at 5.2 AU for $\rho_{\rm s}
 = 1\,{\rm g/cm}^3$ and 
$f_{\rm d} \approx 9.3 \times 10^{-5}$, $3.1\times 10^{-4}$, and $9.3
 \times 10^{-4}$. The thin short line indicates the line proportional to
 $t^3$ for reference. 
\label{fig:mass_evo}
}
\end{figure}

\begin{figure}[htbp]
\epsscale{0.99}
\plotone{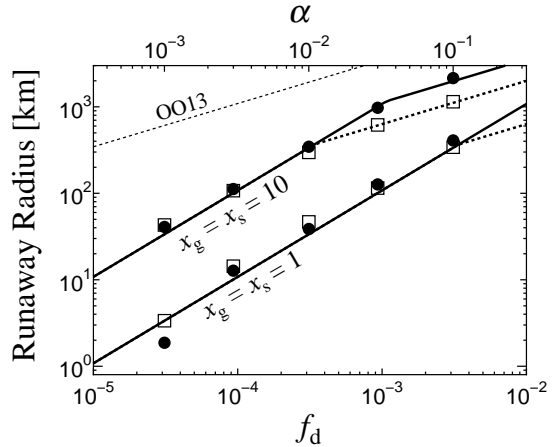}
\figcaption{
Weighted average radius of planetesimals at the onset of runaway growth
 obtained from simulations for $\rho_{\rm s} = 1\,{\rm g/cm}^3$ (open
 squares) and $0.1\,{\rm g/cm}^3$ (filled circles), as
 a function of $f_{\rm d}$ or $\alpha$ for $H_{\rm res,0} = H$. 
The lines correspond to 
the smaller of $r_{\rm p,run,c}$ and $r_{\rm p,run,g}$ given by
 Eqs.~(\ref{rp_col}) and (\ref{eq:rp_gas}) with $\xi_1 = \xi_2 =
 1.5$, respectively, for
 $\rho_{\rm s} = 1\,{\rm g/cm}^3$ (dotted) and $0.1\,{\rm g/cm}^3$
 (solid). The thin dotted line indicates the analytical estimate of the
 runaway radius by \citet{ormel13} for $\rho_{\rm s} = 1\,{\rm g/cm}^3$ in
 10\,MMSN. 
\label{fig:runaway_rudius}
}
\end{figure}

Fig.~\ref{fig:mass_evo} shows the mass evolution of the largest
planetesimals\footnote{The mass of the largest bodies is given by the average mass
of ``runaway bodies'' defined in \citet{kobayashi+10}.}, $M$, in the
radial mesh around 5.2\,AU. The mass evolution with time $t$ is initially proportional
to $M \propto t^3$.  During orderly growth, 
gravitational focusing is less effective; namely, $\dot M$ is
proportional to $M^{2/3}$.  Therefore, the time integration of $\dot M$ results in $M \propto
t^3$. However, the slopes $d \ln M/d \ln t$ become steeper, which 
is explained by the stronger dependence of $\dot M$ on $M$ due to 
gravitational focusing. 
The increase of the slopes \rev{thus} indicates the onset of runaway growth. 
Strong turbulence, i.e., large $f_{\rm d}$, delays the onset of runaway
growth. The onset of runaway growth increases $d \ln M/d \ln t$ to greater than 3. 
The runaway radius, which is defined as the weighted average
radius of planetesimals at $d \ln M/d \ln t = 4$, is obtained
from the simulation in 
Fig.~\ref{fig:runaway_rudius}. The obtained runaway radius is in agreement with
${\bar r}_{\rm p}$ for ${\bar e} \sim v_{\rm esc}/v_{\rm K}$ (see
Figs. \ref{fig:MMSN}, \ref{fig:ecc_radius}, and \ref{fig:runaway_rudius}). 
The runaway radius 
increases with $f_{\rm d}$, so that the runaway
growth is delayed due to strong turbulence.   
 
\begin{figure}[htbp]
\epsscale{0.99}
\plotone{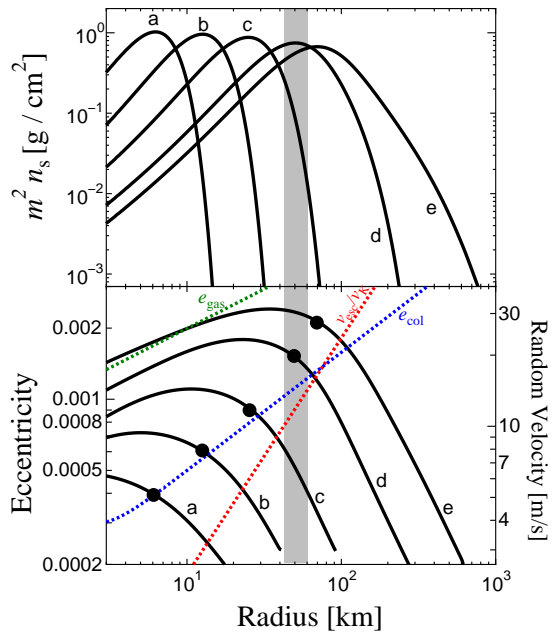}
\figcaption{
Same as Fig.~\ref{fig:MMSN}, but for $\rho_{\rm s} = 0.1 \,{\rm g/cm}^3$ at (a) $1.4\times 10^4$\,yr, (b) $3.6\times 10^4$\,yr, (c)
 $8.9 \times 10^4$\,yr, (d) $2.3 \times 10^5$\,yr, and (e) $3.6 \times 10^5$\,yr. 
\label{fig:MMSN_rho0.1}
}
\end{figure}

For evolution from dust aggregates to pebble or planetesimal sized
bodies, compaction due to the rearrangement of dust aggregates by
self-gravity increases the filling factor of bodies and the internal
density of bodies. However, the filling factor only increases up to $\sim
0.1$ due to self-gravity \citep{kataoka13b}. According to the filling
factor, the internal density of planetesimals is set to be $\rho_{\rm s}
= 0.1\,{\rm g/cm}^3$. Fig.~\ref{fig:MMSN_rho0.1} shows the mass and
velocity evolution of planetesimals for $\rho_{\rm s} = 0.1 \,{\rm
g/cm}^3$ in the MMSN disk with $f_{\rm d} \approx 3.1 \times 10^{-4}$. The
random velocity is smaller than that for high $\rho_{\rm s}$ because
collisional damping and gas drag are more effective. However, 
$v_{\rm esc}$ is smaller for lower $\rho_{\rm s}$, so that runaway growth of
the bodies occurs at 40--50\,km in radius, which is similar to the result
for $\rho_{\rm s} = 1\,{\rm g/cm }^3$.  Interestingly, the radius
\rev{at the onset of} the runaway growth of bodies is almost independent of
$\rho_{\rm s}$ (see also Fig.~\ref{fig:runaway_rudius}).


\section{Analytic Solution for Orderly Growth in a Turbulent Disk}
\label{sc:analy}

The mass distribution of planetesimals depends on the mass growth rate, $\dot
m$. If $p \equiv d \ln \dot m/d \ln m < 1$, smaller bodies grow faster
and the orderly growth of bodies then occurs; bodies with the 
weighted average radius, which approximately determines the solid
surface density, 
is comparable to the radius of the largest bodies.  The collisional
evolution of dust aggregates is an example of orderly growth
\citep[e.g.,][]{okuzumi12}. However, if $p \geq 1$, 
then larger bodies grow faster and hence the runaway growth of bodies
occurs; the weighted average radius becomes much smaller than the radius of
the largest bodies.  

The growth rate $\dot m$ is proportional to the
collisional cross section between bodies with radii $r_1$ and $r_2$,
given by $\pi (r_1 + r_2)^2 (1+ \Theta)$, where $\Theta = v_{\rm esc}^2/
v_{\rm r}^2$ is the Safronov parameter. 
When $v_{\rm r}$ is much smaller than $v_{\rm esc}$ ($\Theta \gg 1$), $\dot m
\propto m^{4/3} v_{\rm r}^{-2}$. The collisional growth phase of bodies is
determined by the mass dependence of $v_{\rm r}$.  The cross section of
gravitational interaction is proportional to the 90 degree scattering
cross section given by $\pi (r_1 + r_2)^2 \Theta^2$, which is larger
than the collisional cross section for $\Theta \ga 1$. The dynamical
friction caused by gravitational interaction results in the energy
equipartition of bodies. If bodies \rev{mainly} grow through collisions with
similar sized bodies, then the random velocities of bodies are proportional
to $m^{-1/2}$ \rev{due to equipartition}, which results in $\dot m \propto m^{7/3}$. $p$ \rev{is}
then larger than unity \rev{and runway} growth occurs. Therefore, $\Theta$ is very important
to derive the radius of planetesimals \rev{at the onset of} runaway growth, and hence, 
we first derive the random velocity of planetesimals in a turbulent
disk.


The random velocity of planetesimals stirred by the density fluctuation
caused by turbulence \rev{can} exceed the surface escape velocity of the planetesimals
($\Theta < 1$). 
Gravitational focusing and dynamical friction are suppressed for $\Theta
< 1$; therefore, 
orderly growth occurs until 
the relative velocity is comparable to the escape velocity. During orderly
growth, 
\rev{a monodisperse population of planetesimals with mass $m$ and radius
$r_{\rm p}$ can be assumed in each narrow annulus of the disk. }
\citet{ormel13}
analytically derived the random velocity through the equilibrium between 
stirring by turbulence and damping by gas drag and \rev{the radius
of planetesimals at }the onset of
runaway growth, using the equilibrium velocity. However, collisional damping
and growth are also important to determine the random velocity. 
The runaway radii obtained by simulations are thus much smaller than their
estimates (Fig.~\ref{fig:runaway_rudius}). In this
section, we derive the random velocity from turbulent stirring 
with the effect of collisions or with the damping by gas drag to derive
\rev{the radius of planetesimals at} 
the onset of runaway growth. 


\subsection{Random Velocity}
\label{sc:velocity}

Damping by mutual collisions decreases eccentricity. 
The eccentricity $e_{12}$ of the body at the barycenter of colliding planetesimals with masses
$m_1$, $m_{2}$ and eccentricities $e_1$, $e_2$ is given by
\citep{ohtsuki92} 
\begin{equation}
 e_{12}^2 = \left(\frac{m_1}{m_1+m_2} \right)^2 e_1^2 
  + \left(\frac{m_2}{m_1+m_2} \right)^2 e_2^2, 
\end{equation}
which is derived under the assumption of the random distribution of the
longitudes of perihelion and the ascending nodes. 
Collisions between planetesimals with similar masses are dominant prior
to runaway growth; therefore, a single merging collision decreases the square of
eccentricity by a factor of 1/2. 

The surface number density of planetesimals is $\Sigma_{\rm s}/m$, the
collisional cross section is \rev{approximated} to $\pi r_{\rm p}^2$
because $v_{\rm r} \gg v_{\rm esc}$, the relative velocity is
approximated as $e a \Omega$ because $e \gg i$, and the ``scale
height'' of planetesimals is given by $i a$; therefore, the collisional rate is
approximately given by $(\Sigma_{\rm s}/m) \pi r_{\rm p}^2 (e/i)
\Omega$.  A more accurate collisional rate between planetesimals is
given by $(\Sigma_{\rm s}/m) h_{m,m}^2 a^2 \langle {\cal P}_{\rm col}(m,m)
\rangle \Omega$ (cf. Eq.~\ref{eq:coag}). If $e v_{\rm K} \gg v_{\rm
esc}$ and $e \gg i$, then $h_{m,m}^2 a^2 \langle {\cal P}_{\rm col}(m,m)
\rangle \approx
C_{\rm col} r_{\rm p}^2 e/i$, where $C_{\rm col}
=4.8$ \citep{inaba01}. The error of the crude estimate for the collisional
rate is a factor $C_{\rm col}/\pi \approx 1.5$. 
\rev{The collisional rate multiplied by $-e^2/2$ is equal to $d e^2 /d
t$; therefore}, the collisional damping rate for bodies
with mass $m$ and radius $r_{\rm p}$ is given by
\begin{equation}
 \left. \frac{de^2}{dt} \right|_{\rm coll} = - C_{\rm col} \frac{e^3}{2i}
 \frac{r_{\rm p}^2 \Sigma_{\rm s}}{m} \Omega.\label{eq:coll_e} 
\end{equation}
The collisional damping for inclination is obtained in the same way, 
as
\begin{equation}
 \left. \frac{di^2}{dt} \right|_{\rm coll} = -C_{\rm col}\frac{i e}{2}
  \frac{\Sigma_{\rm s} r_{\rm p}^2}{m}  \Omega. 
\label{eq:coll_i}  
\end{equation}

From Eqs.~(\ref{eq:sdf}), \rev{(\ref{eq:di_sdf})}, (\ref{eq:coll_e}), and (\ref{eq:coll_i}), the time derivatives of
$e^2$ and $i^2$ are given by 
\begin{eqnarray} 
 \frac{de^2}{dt} &=& f_{\rm d} \left(\frac{\Sigma_{\rm g}
			      a^2}{M_*}\right)^2 \Omega
- C_{\rm col} \frac{e^3}{2i} \frac{r_{\rm p}^2\Sigma_{\rm s}}{m} \Omega, 
\label{eq:de}
\\
 \frac{di^2}{dt} &=& \epsilon^2 f_{\rm d} \left(\frac{\Sigma_{\rm g}
			      a^2}{M_*}\right)^2 \Omega
- C_{\rm col} \frac{i e}{2} \frac{r_{\rm p}^2\Sigma_{\rm s}}{m} \Omega. 
\label{eq:di}
\end{eqnarray}
From Eqs.~(\ref{eq:de}) and (\ref{eq:di}), 
\begin{equation}
 \frac{d}{dt} \left(\frac{i}{e}\right) = \left(\epsilon^2 -
	       \frac{i^2}{e^2} \right)
 \frac{f_{\rm d}}{2 e i} \left(\frac{\Sigma_{\rm g}
			      a^2}{M_*}\right)^2 \Omega.\label{eq:i_e}  
\end{equation}
Interestingly, the time derivative of $i/e$ is independent of
collisional damping but depends only on the stirring.  In the steady
state for $i/e$, Eq.~(\ref{eq:i_e}) gives $i/e = \epsilon$.

The equilibrium eccentricity obtained from
$de^2/dt = 0$ in Eq.~(\ref{eq:de}) is overestimated \citep[cf.][]{ida}
because the timescale of collisional damping is comparable to that of 
collisional growth. Therefore, the effect of collisional 
growth should be taken into account. 
The collisional growth rate of bodies is given by 
\begin{equation}
 4 \pi r_{\rm p}^2 \rho_{\rm s} \frac{dr_{\rm p}}{dt} = 
  \frac{C_{\rm col}}{\epsilon}
  r_{\rm p}^2 \Sigma_{\rm s} \Omega, 
\label{eq:dr}
\end{equation}
where $i/e = \epsilon$ is applied. 
From Eqs.~(\ref{eq:de}) and (\ref{eq:dr}), $d e^2/ d r_{\rm p}$ is
obtained as 
\begin{equation}
 \frac{d e^2}{d r_{\rm p}} = 
  \frac{4 \pi f_{\rm d} \epsilon}{C_{\rm col} }
  \left(\frac{\Sigma_{\rm g} a^2}{M_*}\right)^2 
  \left(\frac{\rho_{\rm s}}{\Sigma_{\rm s}}\right)
  - \frac{3e^2}{2r_{\rm p}}.\label{eq:de2dr} 
\end{equation}

From the integration of Eq.~(\ref{eq:de2dr}) over $r_{\rm p}$, 
the collision-turbulence dominated eccentricity, $e_{\rm col}$, is given by 
\begin{eqnarray}
 e_{\rm col}^2 &=& 
  \frac{8 \pi \epsilon f_{\rm d} \rho_{\rm s} r_{\rm p} }{5 C_{\rm
  col} \Sigma_{\rm s}} 
   \left(\frac{\Sigma_{\rm g} a^2}{M_*}\right)^2   
  \left[ 1 - \left(\frac{r_{\rm p}}{r_{\rm p0}}\right)^{-5/2}\right] 
  \nonumber\label{eq:ecol}
  \\ & & 
  + e_0^2\left(\frac{r_{\rm p}}{r_{\rm p0}}\right)^{-3/2},\label{eq:ecol} 
\end{eqnarray}
where $r_{\rm p0}$ and $e_0$ are the initial radius and
eccentricity, respectively. If $r_{\rm p} \gg r_{\rm p0}$, then
\begin{eqnarray}
e_{\rm col} &\approx& 
  \left(\frac{8 \pi f_{\rm d} \epsilon \rho_{\rm s} r_{\rm
  p}}{5 C_{\rm col} \Sigma_{\rm s} } \right)^{1/2}
   \left(\frac{\Sigma_{\rm g} a^2}{M_*}\right)  
\nonumber
\\
&\approx& 1.5 \times 10^{-4} \left(\frac{f_{\rm d}}{10^{-4}}\right)^{1/2}
 \left(\frac{r_{\rm p}}{1\,{\rm km}}\right)^{1/2}
 \left(\frac{\Sigma_{\rm g}}{140 \, {\rm g \, cm}^{-2}}\right) 
\nonumber 
\\
&&\times 
 \left(\frac{\Sigma_{\rm s}}{2.5 \,{\rm g\,cm}^{-2}}\right)^{-1/2}
 \left(\frac{\rho_{\rm s}}{0.1\,{\rm g/cm}^3}\right)^{1/2}. 
\end{eqnarray}
\rev{The equilibrium eccentricity 
easily estimated from $de^2/dt = 0$ in Eq.~(\ref{eq:de}) with $i/e =
\epsilon$ 
}
is overestimated by a
factor of $\sqrt{5/3}$. 

On the other hand, for large planetesimals and/or small $\Sigma_{\rm s}$, the
eccentricity damping by gas drag is more effective than the collisional
effects. The turbulent stirring given by Eq.~(\ref{eq:sdf}) increases
$e$ effectively, so that $v_{\rm r} = e v_{\rm K}$ is much greater than
$i v_{\rm K}$ and the velocity difference between $v_{\rm K}$ and the
rotational velocity of gas. 
The stopping timescale by gas drag is $\sim (m/ \pi r_{\rm p}^2) (H/\Sigma_{\rm g}) (1/ e a \Omega)$ \citep{adachi76}. The $e^2$-damping rate via gas
drag is approximately given by $e^2$ divided by the stopping time. 
The $e^2$-damping rate is given by 
\begin{equation}
  \left. \frac{de^2}{dt} \right|_{\rm gas} = - 
   B \frac{\Sigma_{\rm g}}{\rho_{\rm s} r_{\rm p}} \frac{a}{H} e^3
   \Omega,\label{eq:gas} 
\end{equation}
where $B \simeq 0.378$ is the correction factor obtained from accurate
treatment of $e$-damping and orbital averaging
\citep{adachi76}.
The equilibrium
eccentricity between gas damping and turbulent stirring is
obtained from Eqs.~(\ref{eq:sdf}) and (\ref{eq:gas}), given
by 
\begin{eqnarray}
 e_{\rm gas} &=& 
\left(\frac{f_{\rm d} \rho_{\rm s} r_{\rm p}
\Sigma_{\rm g} a^3 H }
{B M_*^2} \right)^{1/3} 
\label{eq:egas} 
\nonumber
\\
&\approx& 6.3 \times 10^{-4} \left(\frac{f_{\rm d}}{10^{-4}}\right)^{1/3}
 \left(\frac{r_{\rm p}}{1\,{\rm km}}\right)^{1/3}
 \left(\frac{\Sigma_{\rm g}}{140 \, {\rm g \, cm}^{-2}}\right)^{1/3}
\nonumber 
\\ 
&& \times 
 \left(\frac{a}{5.2 \, {\rm AU}}\right)^{4/3}
 \left(\frac{H}{0.071a}\right)^{1/3}  \left(\frac{\rho_{\rm s}}{0.1\,{\rm g/cm}^3}\right)^{1/3}.
\end{eqnarray}
If $r_{\rm p} \ga
6000\,$km (600\,km) for $\rho_{\rm s} = 0.1\,{\rm g/cm}^3$ ($1\,{\rm
g/cm}^3$) in the MMSN disk with $f_{\rm d} = 10^{-4}$, then $e_{\rm gas} \la e_{\rm
col}$. 
The damping by gas drag is thus less effective than collisions unless planetesimals
are very large. 

The smaller of analytic solutions, $e_{\rm col}$ and $e_{\rm gas}$,
reproduces the evolution curves given by simulations unless the runaway
growth of planetesimals begins (see Fig.~\ref{fig:ecc_radius}).  After the
runaway growth, $e$ of the representative planetesimals is mainly controlled by
stirring due to the largest planetesimals, so that the analytic solutions are no
longer valid. 

\subsection{Runaway Radius}
\label{sc:runaway}

Once $e v_{\rm K} \sim v_{\rm esc}$, runaway growth starts due to
gravitational focusing and dynamical friction. For $e \approx e_{\rm col}$, 
the radius of planetesimals \rev{at the onset of} runway growth, $r_{\rm p,run,c}$,
is determined by $e_{\rm col} = \xi_1 v_{\rm esc} /v_{\rm K}$ with
$\xi_1 \sim 1$. 
Using Eq.~(\ref{eq:ecol}), $r_{\rm p,run,c}$  is given by 
\begin{eqnarray}
 r_{\rm p,run,c} &=& 
  \frac{3 \epsilon f_{\rm d} \Sigma_{\rm g}^2 a^3}{5 \xi_1^2 C_{\rm col} \Sigma_{\rm s} M_*},\label{rp_col} 
\nonumber
\\
 &\approx& 100 \left(\frac{\xi_1}{1.5}\right)^{-2}  
\left(\frac{f_{\rm d} }{10^{-3}}\right)
  \left(\frac{a}{5.2\,{\rm AU}}\right)^{3} 
  \left(\frac{\Sigma_{\rm s}}{2.5\,{\rm g/cm}^2}\right)^{-1} 
\nonumber
\\
&&
 \times 
 \left(\frac{\Sigma_{\rm g}}{140\,{\rm g/cm}^2}\right)^{2} 
  \left(\frac{M_*}{M_\sun}\right)^{-1} 
\, {\rm km}, 
\end{eqnarray}
where $r_{\rm p,run,c} \gg r_{\rm p0}$ is assumed. 
Interestingly $r_{\rm p,run,c}$ is independent of $\rho_{\rm
s}$, which is in 
agreement with the simulation results (see Fig.~\ref{fig:runaway_rudius}). 

If $e$ is determined by $e_{\rm gas}$, 
then the radius of planetesimals
\rev{at the onset of} runaway growth, $r_{\rm p,run,g}$, is obtained 
from $e_{\rm gas} = \xi_2 v_{\rm esc} /v_{\rm K}$ as 
\begin{eqnarray}
 r_{\rm p,run,g} &=& \xi_2^{-3/2} 
\left(\frac{27 f_{\rm d}^2 \Sigma_{\rm g}^2 H^2 a^3}{512 \pi^3 B^2 \rho_{\rm
 s} M_*}\right)^{1/4}
\label{eq:rp_gas} 
\nonumber
\\
&\approx& 350 \left(\frac{\xi_2}{1.5}\right)^{-3/2} \left(\frac{f_{\rm d}}{10^{-3}}\right)^{1/2}
 \left(\frac{H}{0.071a}\right)^{1/2}
 \nonumber
 \\
&&
 \times 
 \left(\frac{a}{5.2\,{\rm AU}}\right)^{3/4} 
 \left(\frac{\Sigma_{\rm g}}{140\,{\rm g/cm}^2}\right)^{1/2}
\nonumber
\\
 &&\times 
 \left(\frac{\rho_{\rm s}}{0.1\,{\rm g/cm}^3}\right)^{-1/4}
\left(\frac{M_*}{M_\sun}\right)^{-1/4} 
\,{\rm km}. 
\end{eqnarray}
\citet{ormel13} derived a formula similar to
Eq.~(\ref{eq:rp_gas}) from the equilibrium between the turbulence stirring and gas drag
damping. 
The formula derived by \citet{ormel13} overestimates by a factor of approximately 3.1 when 
compared to Eq.~(\ref{eq:rp_gas}) with the realistic value of $\xi_2=1.5$
(see also Fig.~\ref{fig:runaway_rudius}). 

The runaway radii given by
simulations are consistent with those obtained from the analytic formulae. Assuming
$\xi_1 = \xi_2 = 1.5$, the smaller of $r_{\rm p, run,c}$ and
$r_{\rm p, run,g}$ reproduces the radius of planetesimals \rev{at the
onset of} runaway
growth, $r_{\rm p,run}$, obtained from the simulations (see
Fig.~\ref{fig:runaway_rudius}).  The runaway radius is determined by $r_{\rm
p,run,c}$ for small $f_{\rm d}$, while
$r_{\rm p,run} \approx r_{\rm p,run,g}$ for large $f_{\rm d}$. 
Strong turbulence (high $f_{\rm d}$) causes $r_{\rm
p,run}$ to be large. This dependence is explained by Eqs.~(\ref{rp_col}) and
(\ref{eq:rp_gas}).
If $r_{\rm p,run} \la 300\,$km around 5\,AU, then $r_{\rm p,run}$ is determined by
$r_{\rm p, run,c}$; therefore, the effects of collisions are more important to
determine the runaway radius. 

\section{Subsequent growth}
\label{sc:subsequent}

The runaway radius, which is derived in \S \ref{sc:runaway} as a
function of turbulence strength, is the typical radius of planetesimals
in the oligarchic growth of planetary embryos subsequent to runaway
growth. Thus, the runaway radius has a strong effect on the growth of embryos,
or the formation of planets.  \citet{kobayashi+11} conducted
simulations for the formation and growth of planetary embryos through
collisional merge and fragmentation, taking into account enhancement
of the collisional cross section by a thin atmosphere, which is mainly important
for Mars-sized or larger planetary embryos. \citet{kobayashi+11} did not
include turbulent stirring, but instead set the initial planetesimal radius
as a parameter. Runaway growth occurs from the beginning in their
simulations. Therefore, the initial planetesimal radii that were set almost
correspond to the runaway radii obtained in the present work.  For
kilometer-sized or larger planetesimals, fragments caused by collisional
shattering are re-accumulated by self-gravity, which determines the
effective collisional strength of the planetesimals.  Small planetesimals
are effectively brittle due to low self-gravity. The growth
of planetary embryos via collisions with small initial 
planetesimals easily stalls due to a reduction of the 
planetesimals caused by active collisional fragmentation and the rapid
radial drift of the yielded fragments. The resultant embryos depend on
the typical planetesimal size during oligarchic growth of the embryos;
If small planetesimals start runaway growth, planetary embryos grow via
collisions with small planetesimals and the collisional fragmentation of
the small planetesimals is too effective to form massive embryos.  On
the other hand, larger representative planetesimals \rev{at the onset of} runaway
growth require a longer timescale for the formation of embryos.  Therefore, we
expect that strong turbulence results in large runaway radii and
produces massive planetary embryos, whereas small runaway radii caused
by weak turbulence is required for the rapid formation of planets.

When the probable planetesimal sizes for the masses and formation
timescales of planets are determined, constraints can then be conversely given on
the strength of turbulence. 
The turbulence strength $f_{\rm d,run}$
required for the runaway radius $r_{\rm p,run}$ is obtained from 
Eq.~(\ref{rp_col}) or (\ref{eq:rp_gas}). 
Under the assumption of the power-law disk of
Eqs.~(\ref{eq:sigmag}) and (\ref{eq:sigmas}), $f_{\rm d,run}$ is given by the larger of $f_{\rm d,run,c}$ or 
$f_{\rm d,run,g}$, where
\begin{eqnarray}
 f_{\rm d,run,c} &=& 9.7 \times 10^{-4} 
\frac{f_{\rm ice} x_{\rm s}}{x_{\rm g}^2}
\left(\frac{r_{\rm p}}{100\,{\rm km}}\right)
\left(\frac{a}{5.2\,{\rm AU}}\right)^{-3/2} 
\nonumber\label{eq:fdc}
\\&&\times 
\left(\frac{M_*}{M_\sun}\right),
\\
 f_{\rm d,run,g} &=& 6.7 \times 10^{-9} x_{\rm g}
\left(\frac{r_{\rm p}}{100\,{\rm km}}\right)^{2}
\left(\frac{H}{0.37\,{\rm AU}}\right)^{-1}
\nonumber\label{eq:fdg}
\\
&& \times 
\left(\frac{\rho_{\rm s}}{0.1\,{\rm g/cm}^3}\right)^{1/2}
\left(\frac{M_*}{M_\sun}\right)^{1/2},
\end{eqnarray}
are derived from Eqs.~(\ref{rp_col}) and (\ref{eq:rp_gas}),
respectively. Because the collisional effect is more important than
damping by gas drag, $f_{\rm d,run}$ is determined by $f_{\rm
d,run,c}$. Constraints on $f_{\rm d}$ are given by calculating $f_{\rm
d,run}$ below. 

For the formation of Jupiter, a core should grow up to approximately 10
Earth masses within the disk lifetime and induce rapid gas accretion 
\citep{ikoma00}. 
\rev{
The lifetimes of disks are inferred from the thermal emission of dust as
$\sim 1$--6\,Myr \citep{briceno,haisch,najita}. In the simulation by
\citet{kobayashi+11}, a $10M_\oplus$ core can be formed from 100\,km radius
planetesimals at 5\,AU in the 10 MMSN disk within $1\,$Myr 
}\footnote{\rev{
In their result for 100\,km radius initial planetesimals in the 10
MMSN (Fig.~5 of \citet{kobayashi+11}), cores are larger than
$10\,M_\oplus$ inside 10\,AU at 10\,Myr. The core formation timescale at
5\,AU is approximately 10 times shorter than that at 10\,AU; therefore, a $10 M_\oplus$ core
at 5\,AU can be formed within $1\,$Myr. The data of the simulation conducted by
\citet{kobayashi+11} do show the formation of $10M_\oplus$ cores within $1\,$Myr. 
}}\rev{.}
From Eqs.~(\ref{eq:fdc}) and (\ref{eq:fdg}), $f_{\rm d,run} \sim
10^{-4}$ is required for $r_{\rm p,run} \sim 100$\,km at 5\,AU in
10\,MMSN. Fig.~\ref{fig:runaway_rudius} 
shows $f_{\rm d}$ 
(or $\alpha$ for $H_{\rm res,0} \approx H$), which can prepare the conditions
for the formation of Jupiter given by $r_{\rm p,run} = 30$--300\,km at
5--6\,AU in disks with $x_{\rm g} = x_{\rm s} = 10$. 

\begin{figure}[htbp]
\epsscale{2} 
\centering 
\plottwo{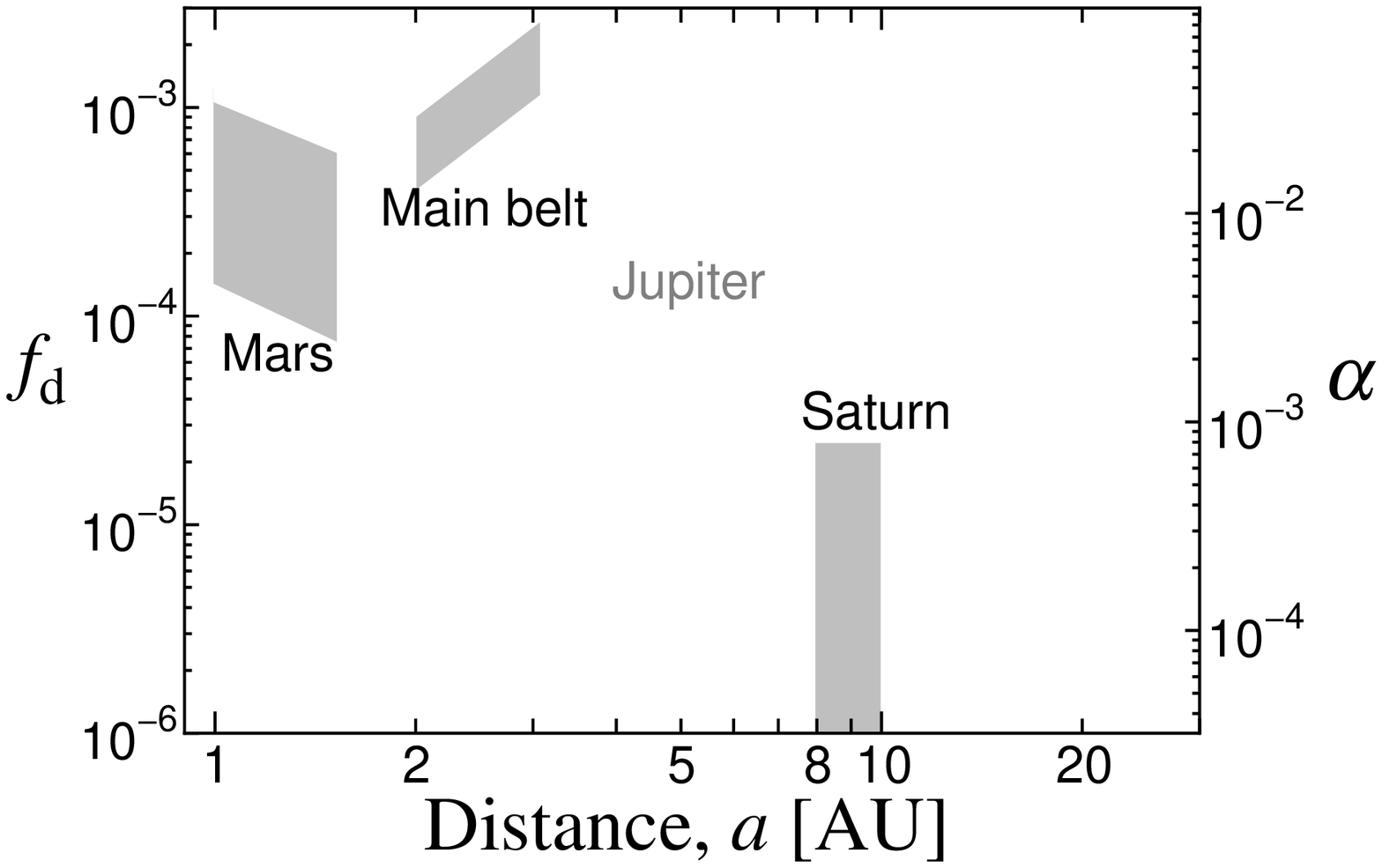}{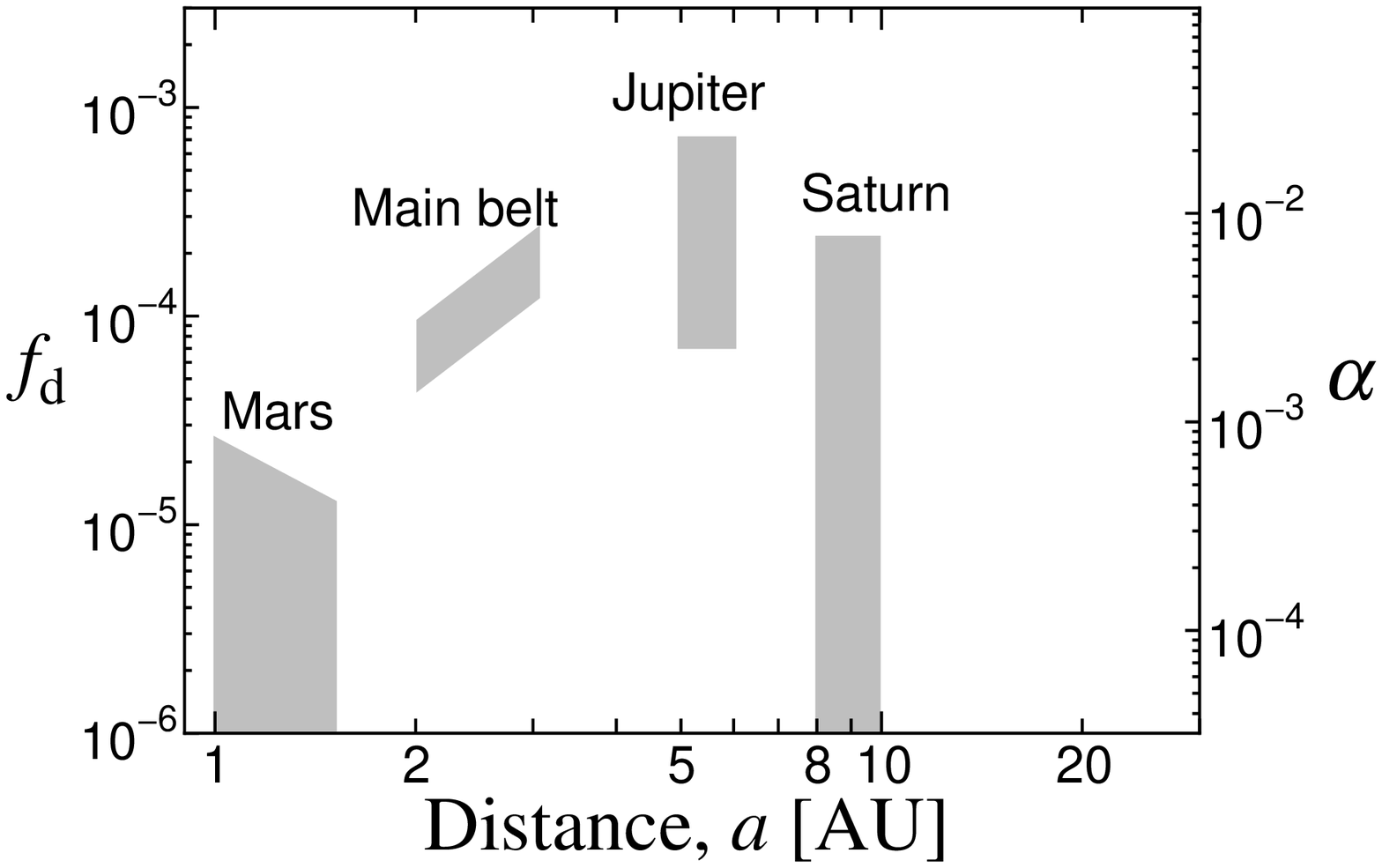} \figcaption{
Turbulence strength $f_{\rm d}$ or $\alpha$, required for the formation
of Mars, Jupiter, and Saturn for the MMSN (top panel) and 10~MMSN (bottom
panel).  The formation of the $\sim 10 M_\oplus$ core required for Jupiter via
 core accretion is difficult to accomplish with the MMSN. 
The value of $\alpha$ is estimated under the assumption
of $H_{\rm res,0} = H$.  
\label{fig:semi_runaway}
} 
\end{figure}

Saturn's core forms in the disk after the formation of Jupiter.  The gap
opening by Jupiter assists the rapid growth of Saturn's core at the outer
edge of the gap, where the radial drift due to gas drag is
negligible. The collisional fragments produced during core formation
accelerate the growth of the core, and the supply of fragments produced
in the outer disk induces further rapid growth. These effects allow
the rapid formation of Saturn's core within $\sim 10^6$ years after gap
opening by Jupiter, and the rapid formation may explain the massive core
of Saturn \citep{kobayashi12}.  For the rapid growth of Saturn's core in
10 MMSN,
$100$\,km radius or smaller planetesimals are required, which is satisfied for
$f_{\rm d} \la 3 \times 10^{-4}$ ($\alpha \la 8 \times 10^{-3}$ for
$H_{\rm res,0} = H$) in the 10~MMSN disk (see
Fig.~\ref{fig:semi_runaway}). This upper limit is similar to the
condition for Jupiter. If Jupiter can be formed via other mechanisms in
the MMSN disk, then kilometer sized or smaller
planetesimals are required for the rapid formation of Saturn's core,
which corresponds to $\alpha \la 10^{-3}$ for $H_{\rm res,0} = H$. 

On the other hand, the formation age of Mars estimated from the amount
of W isotope in the mantle is 2--4\,Myr \citep{dauphas}. For 10\,MMSN (MMSN), the early formation of
Mars may be explained by planetesimals with radii of $\la 10\,$km (5--30\,km) 
\citep{kobayashi13}.  
The disk inside the snow line ($a < 2.7$\,AU) has $f_{\rm ice} \approx
0.24$ \rev{\citep{hayashi}}.
From Eqs.~(\ref{eq:fdc}) and (\ref{eq:fdg}), the formation of Mars requires 
$f_{\rm d} < 2 \times 10^{-5}$ at 1\,AU 
($f_{\rm d} < 1 \times 10^{-5}$ at 1.5\,AU)
in the 10~MMSN disk, while $f_{\rm d} \approx 10^{-4}-10^{-3}$ is
suitable at 1--1.5\,AU in the MMSN (see Fig.~\ref{fig:semi_runaway}).  

In addition, main-belt asteroids with radii smaller than
50--100\,km have a $d \ln n_{\rm s}/d \ln m$ slope of
$\sim -11/6$, which is similar to the outcome of collisional cascade \citep{bottke}, while
larger main-belt asteroids except for the largest bodies (Ceres, Vesta, and
Pallas) have shallower slopes. The main-belt asteroids, except for the
largest asteroids, have a peak of $m^2 n_{\rm s}$ in the radius range of 50--100\,km. The onset of runaway growth causes a similar shaped peak around the runaway radius 
(see Figs.~\ref{fig:MMSN} and \ref{fig:MMSN_rho0.1}). 
Assuming that the radius of 50--100\,km in the main belt 
corresponds to the planetesimal radius, $f_{\rm d} = (5 \dots 9)
\times 10^{-5}$ at 2\,AU ($f_{\rm d} = (1 \dots 3) \times 10^{-5}$
at 3\,AU) in the 10 MMSN and $f_{\rm d} = (4 \dots 10)
\times 10^{-4}$ at 2\,AU ($f_{\rm d} = (1 \dots 3) \times 10^{-3}$
at 3\,AU) in the MMSN. 

In Fig.~\ref{fig:semi_runaway}, the constraints on the
turbulence strength ($f_{\rm d}$ or $\alpha$) are summarized based on these considerations
.  
Jupiter cannot be formed via planetesimal accretion in the MMSN; therefore, 
there is no hatched region for Jupiter in the MMSN. 
If the Solar System was formed in a disk as massive as the 10\,MMSN disk,
then the radial distribution of turbulence is that shown in the bottom panel of
Fig.~\ref{fig:semi_runaway}. 
In the inner disk where terrestrial planets formed,
turbulence is expected to have been weak, and thus the disk may be a
``dead zone'' where MRI is substantially suppressed.  On the other hand,
the disk beyond 2\,AU is expected to have had relatively high $f_{\rm
d}$ or $\alpha$. 
\rev{Note that the radial dependence of $f_{\rm d}$ or $\alpha$ 
also depends on that of
$\Sigma_{\rm g}$ and $\Sigma_{\rm s}$. If the surface density
for solid and gas is comparable to that of 10\,MMSN around 5\,AU, as required for the
formation of Jupiter's core, then $\alpha$ required for the formation of Mars 
in the disk with $\Sigma_{\rm g} \propto \Sigma_{\rm
s} \propto a^{-1}$ is
approximately 3 times larger than that estimated in Fig.~\ref{fig:semi_runaway}.}

\rev{
In MRI, $\alpha$ is a dimensionless coefficient that is proportional to 
the turbulent accretion stress integrated over the elevation of a disk. 
The saturation value of $\alpha$ in MRI simulations depends on 
conditions such as the initial vertical net magnetic field $\langle B_{z,0} \rangle$,
and the ohmic resistivity $\eta$. 
Based on MRI simulations, 
\citet{okuzumi11} empirically derived 
\begin{equation}
 \alpha = \frac{510}{\beta_{z,0}} \exp\left(\frac{0.54 H_{\rm
				      res,0}}{H}\right) 
 + 0.0011 \exp \left(\frac{3.6 H_{\Lambda,0}}{H}\right), 
\end{equation}
where $\beta_{z,0} = 8\pi \rho_{\rm gas} \langle B_{z,0} \rangle$ is the plasma
beta and $H_{\Lambda,0}$ is the vertical width of the dead zone that 
includes the resistive MRI zone. 
The charge reaction network gives the scale heights of the dead zones,
$H_{\rm res,0}$ and $H_{\Lambda,0}$. However, the results depend on the
total surface area of solid particles \citep[e.g.,][]{ilgner06}. 
In addition, $\alpha$ depends on $\langle B_{z,0} \rangle$, which are determined by the
radial transport of large-scale magnetic fields in accretion disks
\citep{takeuchi}.  
Therefore, the $\alpha$ value is uncertain due to conditions such as 
the magnetic fields and the
total surface area of particles. 
However, the results from attempts at simulation in this work may provide some constraints on these physical
conditions during planet formation. 
}

\section{Summary and Discussion}
\label{sc:discussion}

Turbulent stirring increases the relative velocities between planetesimals and thus
delays the onset of runaway growth; instead orderly growth occurs,
in which the solid surface density or the total mass of planetesimals is mainly
determined by planetesimals with the weighted average radius.  The
relative velocity between the representative planetesimals, $v_{\rm r}$,
is much higher than their surface escape velocity, $v_{\rm esc}$, in the early stage. The
representative planetesimals get larger due to collisional growth, so that the Safronov
parameter $\Theta = (v_{\rm r}/v_{\rm esc})^2$ is smaller and the
runaway growth of planetesimals then starts when $\Theta \sim 1$. This has been confirmed
via simulations described in Section \ref{sc:simulation}. In addition,
we have analytically derived the following solutions, which perfectly
reproduce the results of the simulations. 
\begin{enumerate}
 \item When the representative planetesimals are small or 
the solid surface density is large, the random velocity is determined by
the collisional \rev{effects (damping and coagulation of planetesimals)} and turbulent stirring,
which is given by Eq.~(\ref{eq:ecol}). 
 \item For large planetesimals and/or low 
solid surface density, the random velocity is given by the equilibrium
between gas damping and turbulent stirring, as given by
Eq.~(\ref{eq:egas}). 
 \item When $\Theta \approx 2.25$ or $v_{\rm r} \approx 1.5 v_{\rm
       esc}$, the runaway growth of bodies begins.  Using the random
       velocities given in Eqs.~(\ref{eq:ecol}) and (\ref{eq:egas}), the
       radius of planetesimals \rev{at the onset of} runaway growth, $r_{\rm p, run}$, is
       determined by the smaller of Eqs.~(\ref{rp_col}) or
       (\ref{eq:rp_gas}). For the internal density $\rho_{\rm s} =
       0.1\,{\rm g/cm}^3$ ($1\,{\rm g/cm}^3$), if $r_{\rm p,run} \la
       400\,$km (200\,km), then the collisional effect is more important than
       the gas drag, and thus $r_{\rm p,run}$ is given by
       Eq.~(\ref{rp_col}).
\end{enumerate}

Subsequent growth is strongly affected by the radius of planetesimals
\rev{at the onset of} runaway growth $r_{\rm p,run}$ because their typical size is
almost unchanged during subsequent growth. Taking into account the
previous studies on planet formation starting from \rev{initially
large} planetesimals \rev{in non-turbulent disks}, the following has been determined:
\begin{enumerate}
 \item For the formation of Jupiter via the core accretion scenario, $\sim 10
       M_\oplus$ is formed within the disk lifetime, which requires
       $\sim 30$--300\,km planetesimals in a massive disk $x_{\rm g} \sim
       x_{\rm s} \sim 10$ \citep{kobayashi+11}. The turbulence strength
       with $f_{\rm d} \sim 10^{-3}$--$10^{-4}$ ($\alpha \sim 3$--$30
       \times 10^{-3}$ for $H_{\rm res,0} = H$) results in $r_{\rm
       p,run} \sim 30$--300\,km at 5--6\,AU in the 10~MMSN disk, which
       may produce a massive core for the formation of Jupiter.
 \item For the formation of Saturn, the gap opening by Jupiter assists the rapid
       accretion of Saturn's core after the formation of Jupiter; 
       \rev{if representative planetesimals at the onset of runaway
       growth are smaller than 100\,km,} then Saturn's core \rev{can be
       formed} within $10^6$ years after 
       the formation of Jupiter \citep{kobayashi12}.
Therefore, the formation of Saturn requires a turbulence strength $f_{\rm
       d} \la 2\times 10^{-4}$ ($\alpha \la 10^{-2}$ for $H_{\rm res,0}
       = H$), which satisfies the conditions for the formation of Jupiter.
 \item The core formation \rev{timescale} of Mars is estimated to be 2--4 Myrs
       \citep{dauphas}, which requires 10\,km or smaller planetesimals
       at the onset of runaway growth in a massive disk required to form
       Jupiter's core. In the 10~MMSN disk, weaker turbulence with
       $f_{\rm d} \la 10^{-5}$ ($\alpha \la 3 \times 10^{-4}$ for
       $H_{\rm res,0} = H$) results in $r_{\rm p,run} \la 10$\,km around
       1\,AU, which may form Mars rapidly.  The condition for the formation of Mars 
       corresponds to $f_{\rm d} \la 10^{-5}$ ($\alpha \la 3 \times
       10^{-4}$ for $H_{\rm res,0} = H$) in the 10\,MMSN disk.
 \item The proposed fossil feature in the size distribution of main-belt
       asteroids has a radius of approximately 50--100\,km, which is explained
       by turbulence with $f_{\rm d} \sim 10^{-4}$ ($\alpha \sim
       10^{-2}$ for $H_{\rm res,0} = H$) for the 10\,MMSN disk.
 \item The turbulent strength expected for formation of the Solar System is
       summarized in Fig.~\ref{fig:semi_runaway}, assuming the Solar System was
       formed in the 10\,MMSN disk. The inner 
       disk where terrestrial planets formed may have low $\alpha$,
       while $\alpha$ is larger in the outer disk.
\end{enumerate}

\citet{kobayashi14} investigated the formation of the debris disk caused by
planet formation in planetesimal disks. Planetary embryos formed from
planetesimals induce collisional fragmentation of remnant planetesimals,
which can form debris disks after gas depletion. Narrow disks with
100\,km planetesimals around 30\,AU can explain the evolutionary trend
of infrared excesses of debris disks observed at 18 and 70\,$\micron$ by
the Spitzer Space Telescope. \rev{The turbulent strength required for debris
disks is similar to that for the outer solar system. }

\rev{The formation of Jupiter required a massive
disk. However, the surface density of planetesimals 
constructed via the collisional evolution of bodies is different from
the initial solid surface density \citep{okuzumi12}.
In addition, pebbles grow in the outer disk and drift to the inner
disk. The collisional cross sections between pebbles and planetary
embryos are high \citep[e.g.,][]{ormel_klahr10}. 
The growth of embryos may be accelerated due to the accretion of pebbles
that drift from the outer disk \citep{bromley,lambrechts,levison}. 
Therefore, if the initial protoplanetary disk is even less massive than
10\,MMSN, then the formation of Jupiter may be possible. 
}
In future work, we should address continuous collisional
growth from dust to planets in turbulent disks, which may reveal the formation of the debris
disk as well as planet formation.

We thank Neal Turner for his valuable comments on our manuscript. This work was supported by
Grants-in-Aid for Scientific Research (Nos. 26287101, 23103005, 23103004, 15H02065)
from the Ministry of Education, Culture, Sports, Science and Technology (MEXT) of Japan and by the Astrobiology Center Project of the National
Institute of Natural Sciences (NINS) (Grant Number AB271020).

\end{document}